\documentclass[preprint,prb,aps,amsmath,amssymb,color,superscriptaddress]{revtex4}
\usepackage{stmaryrd}
\usepackage{amsmath}
\usepackage{amssymb}
\usepackage{graphicx}
\usepackage{dcolumn}
\usepackage{bm}
\usepackage{multirow}
\usepackage{tabularx}
\usepackage{epsfig}

\begin{document}
\begin{titlepage}
\title{Transition from band insulator to excitonic insulator via alloying Se into Monolayer TiS$_3$: A Computational Study}
\author{Shan Dong}
\affiliation{Key Lab of advanced optoelectronic quantum architecture and measurement (MOE), and Advanced Research Institute of Multidisciplinary Science, Beijing Institute of Technology, Beijing 100081, China}
\author{Yuanchang Li}
\email{yuancli@bit.edu.cn}
\affiliation{Key Lab of advanced optoelectronic quantum architecture and measurement (MOE), and Advanced Research Institute of Multidisciplinary Science, Beijing Institute of Technology, Beijing 100081, China}
\date{\today}

\begin{abstract}
First-principles density functional theory plus Bethe-Salpeter equation calculations are employed to investigate the electronic and excitonic properties of monolayer titanium trichalcogenide alloys TiS$_{3-x}$Se$_x$ ($x$=1 and 2). It is found that bandgap and exciton binding energy display asymmetric dependence on the substitution of Se for S. While the bandgap can be significantly decreased as compared to that of pristine TiS$_3$, the exciton binding energy just varies a little, regardless of position and concentration of the Se substitution. A negative exciton formation energy is found when the central S atoms are replaced by Se atoms, suggesting a many-body ground state with the spontaneous exciton condensation. Our work thus offers a new insight for engineering an excitonic insulator.
\end{abstract}

\maketitle
\draft
\vspace{2mm}
\end{titlepage}

\vspace{0.3cm}
\textbf{I. Introduction}
\vspace{0.3cm}

Excitonic insulator is first proposed in the 1960s by theoretical physicists, which has a many-body ground state with spontaneous exciton formation\cite{Mott,Knox,Keldysh,DesCloizeaux,Kohn,Halperin}. It is a macroscopic quantum state similar to BCS superconductors but resulted from the larger exciton binding energy than the energy gap. The nature of the excitonic insulator phase remains inconclusive although more than fifty years have passed and compelling experimental evidence is still lacking\cite{Kogar}. In recent years, the boom of excitonic insulator research has been reignited, partly by virtue of the vast progress in low-dimensional systems\cite{WangZF,Du,LiZ,Varsano,usEI,usHEI,usgraphone,Varsano2}. For instance, experimental evidences are reported for the topological excitonic insulator in InAs/GaSb quantum well structure\cite{Du} and for the exciton Bose-Einstein condensate at temperature above 100 kelvin in MoSe$_2$-WSe$_2$ atomic double layers\cite{WangZF}. Theoretically, one-dimensional carbon nanotubes\cite{Varsano} and two-dimensional gallium arsenide\cite{usEI}, monolayer transition-metal chalcogenides\cite{usEI,Varsano2} and halides\cite{usHEI}, and semi-hydrogenated graphene\cite{usgraphone} are predicted as the excitonic insulator on the basis of first-principle density functional theory plus Bethe-Salpeter equation (BSE) calculations. The studies in the two-dimensional magnetic systems\cite{usgraphone,usHEI} imply that an exciton instability can even exist in wide-gap semiconductors with the one-electron gap $>$ 3 eV, thus going beyond the conventional understanding that the excitonic insulator transition occurs in small gap semiconductors or semi-metals\cite{Kohn}.

Monolayer TiS$_3$ is one kind of transition metal trichalcogenides, which has been fabricated by exfoliation from its layered material\cite{Island}. It has a direct gap around 1 eV with high carrier mobility and may find applications in future nanoelectronics, thermoelectric device and infrared light emission\cite{Dai,Zhang,Khatibi}. Due to the two-dimensional nature, the screening of electron-hole interaction is significantly reduced in monolayer TiS$_3$ , leading to a large exciton binding energy\cite{usPRL,usEI,Mendoza,Donck,Torun}. Even more interesting is that its band edge states have the same parity and thus transitions between them are dipole forbidden which further suppresses the screening effect. As a consequence, the ground state dark exciton is predicted to have a binding energy just $\sim$0.3 eV smaller than the one-electron bandgap\cite{usEI}. Under a moderate compressive strain, the exciton binding energy would exceed the bandgap, hence realizing the phase transition from a band-insulator to an excitonic-insulator.

Alloying is another widely used method to tailor the electronic and magnetic properties of solid state materials for designing specific applications\cite{Ersan2020,Mahat,Miyazato,Ersan2015,Wang,Braganca,Chen,Liu,Yagmurcukardes,Ersan,Xie,Oshima,Tan}. This has been well-established for transition metal dichalcogenides\cite{Xie,Oshima,Tan}, e.g., MoS$_{2-x}$Se$_x$, Mo$_{1-x}$W$_x$S$_2$, TiS$_{2-x}$Se$_x$ and Nb$_{1-x}$Re$_x$S$_2$, etc. Recently, alloying a fraction of Se into the TiS$_3$ has also been experimentally demonstrated by Agarwal \emph{et al}\cite{Agarwal}. While current success is limited to the alloys with a miniscule amount (8\%) of Se, their energetic calculations on TiS$_{3-3x}$Se$_{3x}$ showed that the stable alloy composition is located at much higher amount $x$ = 1/3. Isoelectronic substitution of S with Se not only can tune the system bandgap as usual but would also generate a so-called chemical pressure which thus may drive the alloy into an intrinsic excitonic insulator state.

In this paper, we investigate the effect of Se-to-S substitution on the electronic and excitonic properties of the TiS$_3$ monolayer, especially the changing trend of the bandgap relative to the exciton binding energy, using the first-principles calculations coupled with the BSE. It is found that the S atoms of the TiS$_3$ located at the middle and the side have substantially different contributions to the band edge states, consequently leading to the very different substitution effects. Substituting middle S with Se significantly reduces the bandgap while substituting side S causes less effect. On the contrary, the introduction of Se almost has no effect on the exciton binding energy no matter what kind of S is substituted, manifesting itself as a unique two-dimensional screening behavior. This disproportionate change between the bandgap and exciton binding energy can eventually cause a phase transition from a band-insulator to an excitonic-insulator as a result of exciton binding energy exceeding the corresponding bandgap when the substitution occurs for the middle S atoms.

\vspace{0.3cm}
\textbf{II. Methodology and models}
\vspace{0.3cm}

Geometric optimizations and electronic structure calculations were performed using the Perdew-Burke-Ernzerhof (PBE) \cite{PBE} exchange-correlation within the framework of density functional theory as implemented in the QUANTUM ESPRESSO code \cite{QE}. The Heyd-Scuseria-Ernzerhof (HSE06) hybrid functional\cite{HSE06} was further used for the bandgap calculations on top of the PBE optimized structures. After a convergence test, a 60 Ry cutoff was set for optimized norm-conserving Vanderbilt pseudopotentials\cite{Hamann}. A vacuum layer more-than-15 \AA\ was used along the out-of-plane direction to avoid spurious interactions between adjacent layers. A $12\times 16\times 1$ $k$-point grid was used for the geometric optimization and all structures were fully relaxed until the residual force on each atom was less than 0.01 eV/\AA. A fine $18\times 24\times 1$ $k$-point grid was adopted for energetic and electronic structure calculations. The BSE was solved for the excitonic properties by using the YAMBO code\cite{yambo}. The same fine $k$-point grid, 88 bands, and 12-Ry cutoff were used to calculate the dielectric function matrix. The top five valence bands and bottom two conduction bands were chosen to build the BSE Hamiltonian. The Coulomb cutoff technique was employed in the BSE calculations to avoid artificial interaction between the periodic adjacent layers. Because of unaffordable computational cost at present for a fully converged solution of BSE at the HSE level, the PBE bandgap is corrected to the HSE values by applying a scissor operator for both the response function and diagonal part of the BSE kernel\cite{Sangalli}. The geometric structures were illustrated with VESTA\cite{VESTA} in the figures.

\vspace{0.3cm}
\textbf{III. Results and discussion}
\vspace{0.3cm}

\begin{figure}[tbp]
\includegraphics[width=0.95\columnwidth]{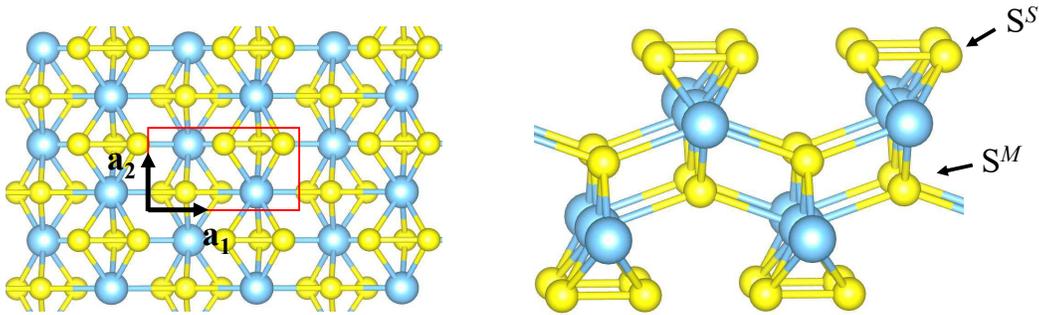}
\caption{\label{fig:fig1} (Color online) (a) Top and (b) side views of monolayer TiS$_3$. Cyan and yellow balls denote Ti and S atoms, respectively. The red rectangle denotes the unit cell with in-plane lattice vectors \textbf{a}$_1$ and \textbf{a}$_2$.}
\end{figure}

Monolayer TiS$_3$ possesses the inversion symmetry and a unit cell is composed of two Ti and six S atoms. It takes a rectangle structure in the basal plane with optimized lattice constant $a_1$=5.017 \AA\ and $a_2$=3.413 \AA, and has a quintuple-layer structure with S-Ti-S-Ti-S atomic layers stacked along the vertical direction, as shown in Fig. 1. There are two kinds of S atom: One is at the side (S$^S$) and the other is at the middle (S$^M$).

Shown in Fig. 2(a) is the band structure of pristine TiS$_3$ monolayer. While the PBE underestimates the bandgap by giving a value of 0.30 eV at the $\Gamma$ point, the HSE increases it to 1.15 eV but holds its direct-gap feature. We also calculate the parity for band edge states as denoted. Being the same odd parity implies that transitions between them are dipole forbidden, which suppresses the system screening and decouples the exciton binding energy from the bandgap. This has been shown to play a key role in the spontaneous formation of exciton condensation\cite{usEI}. Here obtained results are in line with previous studies\cite{Dai,usEI,Kang}.

In order to maintain a well-defined parity, only Se substitutions that can preserve the inversion symmetry are considered throughout the paper. According to where the Se is located, there are three cases, namely, i) two Se replace two side S [denoted as TiS$_2$Se$^S$; see the inset of Fig. 2(b)], ii) four Se replace all four side S (denoted as TiSSe$_2$; see the inset of Fig. 2(c)), and iii) two Se replace two middle S (denoted as TiS$_2$Se$^M$; see the inset of Fig. 2(d)).

\begin{figure}[tbp]
\includegraphics[width=0.95\columnwidth]{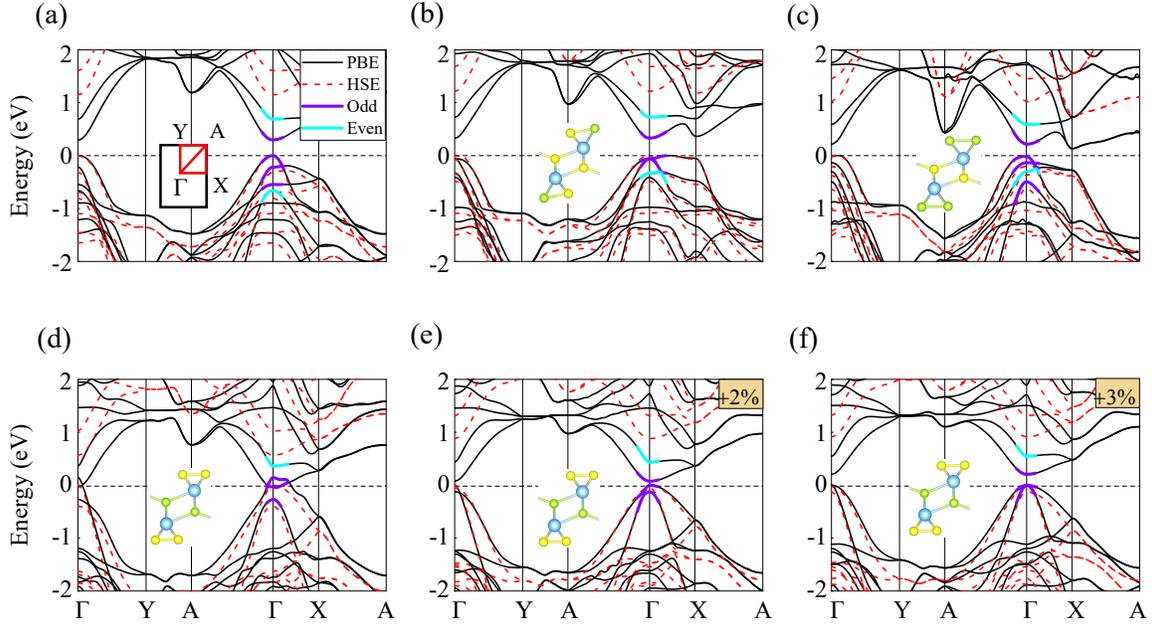}
\caption{\label{fig:fig2} (Color online) Electronic structures of the (a) TiS$_3$, (b) TiS$_2$Se$^S$, (c) TiSSe$_2$, (d) TiS$_2$Se$^M$, (e) TiS$_2$Se$^M$ with 2\%-strain, and (f) TiS$_2$Se$^M$ with 3\%-strain. Black solid lines are calculated by the PBE and red dashed lines are by the HSE06. Band parities are also denoted for the sates close to the Fermi level which is set to energy zero. Shown in the inset of (a) denotes the Brillouin zone for band structure calculations. Shown in the insets of (b)-(f) are the corresponding geometric configurations. Cyan, yellow and orange balls are Ti, S and Se atoms. }
\end{figure}

Their corresponding electronic structures are plotted in Figs. 2(b)-2(d). First of all, we find that the parity of band edge states keeps unchanged regardless of where the Se appears. At the PBE level, both the  TiS$_2$Se$^S$ and the TiSSe$_2$ are characterized by an indirect gap with the values of 0.33 eV and 0.13 eV, respectively. But the TiS$_2$Se$^M$ is predicted to be metallic with a small overlap between the bottom conduction band and top valence band. Actually, the PBE significantly underestimates the bandgap of bulk and monolayer TiS$_3$\cite{Dai}, which should be also the case for the alloys. Previous works showed that the HSE can reproduce the experiment gap for the TiS$_3$\cite{Island,Dai}. Herein, the HSE increases the minimum gap to 1.18 eV, 0.76 eV and 0.58 eV for the TiS$_2$Se$^S$, the TiSSe$_2$ and the TiS$_2$Se$^M$. The former two are still indirect, which are, respectively, 0.03 eV and 0.25 eV smaller than the direct gap at the $\Gamma$ point while the latter is direct at the $\Gamma$ point.

\begin{table*}
\centering
\caption{Comparison of lattice constant, minimum gap, direct gap at the $\Gamma$ point, formation energy ($E_f$) and corresponding binding energy ($E_b$) of the ground-state $X_1$ exciton, as well as two-dimensional polarizability ($\alpha_{2D}$), for the TiS$_{3-x}$Se$_x$. Note that the $E_f$ and $E_b$ for the TiS$_2$Se$^M$ are obtained on the electronic structure under 2\%-strain but correcting the bandgap by a scissor operator to the HSE06 value without strain, namely, 0.58 eV.}
\renewcommand\arraystretch{1.10}
\begin{ruledtabular}
\begin{tabular}{lcccccccccccccccccccccccccc}
   & \multicolumn{2}{c}{lattice constant (\AA)} & \multicolumn{2}{c}{minimum gap (eV)} & \multicolumn{2}{c}{direct gap (eV)} & \multirow{2}{*}{$E_f$ (eV)} & \multirow{2}{*}{$E_b$ (eV)}& \multirow{2}{*}{$\alpha_{2D}$ (\AA)}\\
  \cline{2-3}
  \cline{4-5}
  \cline{6-7}
   & $a_1$ &$a_2$ & PBE & HSE06 & PBE& HSE06 &  \\
\hline
   TiS$_3$ & 5.017 & 3.413 & 0.30 & 1.15 & 0.30 & 1.15 & 0.40 & 0.75 & 5.66\\
      TiS$_2$Se$^S$ & 5.018 & 3.465 & 0.33 & 1.18 & 0.39 & 1.21  & 0.48 & 0.73 & 6.44\\
   TiSSe$_2$ & 5.232 & 3.495 & 0.13 & 0.76 & 0.21 & 1.01 & 0.20 & 0.81 & 6.74\\
   TiS$_2$Se$^M$ & 5.188 &3.455 & metal & 0.58 & metal & 0.58 & -0.13 & 0.71 & 6.13\\
\end{tabular}
\end{ruledtabular}
\end{table*}

These results indicate that the change of bandgap strongly depends on the Se substitution position. Comparison between Figs. 2(a) and 2(b) reveals that when only part of the S$^S$ is replaced, the gap is barely affected and the change of band structure mostly occurs at the $X$ point. But further increasing the Se concentration would lead to the notable minimum-gap reduction due to the gradual down-shift of conduction band at the $X$ point. The TiSSe$_2$ is a case in which all the S$^S$ is replaced by Se, and the HSE gap is decreased by 0.39 eV relative to that of the TiS$_3$. For comparison, the change of direct gap is not so obvious. In contrast, replacing the S$^M$ significantly reduces the HSE bandgap to only about half that of TiS$_3$ (see Table I). The two categories of gap reduction above should be associated to different physical origins\cite{Ersan}.

Both our calculations and previous works\cite{Zhang,Kang} show that the top valence band of monolayer TiS$_3$ is dominantly contributed from the S$^M$. So when these S$^M$ atoms are replaced by the less electronegative Se atoms, the valence band moves upwards, certainly giving rise to a smaller bandgap for the TiS$_2$Se$^M$. Likewise, replacing a fractional of S$^S$ atoms hardly affects the band edge state, so the bandgap is basically unchanged. This is the case of the TiS$_2$Se$^S$. But on the other hand, the Se has a larger atomic radius than that of the S. If too many Se atoms appear at the side, there is not enough space to allow their sufficient relaxation, hence introducing an internal pressure. This is another alternative mechanism that can lead to the gap reduction\cite{usEI}. Some hints might be unveiled from the change of lattice constant as compared in Table I. The TiS$_2$Se$^S$ has the almost same $a_1$ as the TiS$_3$ and the $a_2$ is slightly increased by $\sim$1.5\%. But from the TiS$_2$Se$^S$ to the TiSSe$_2$, $a_1$ is remarkably increased by $\sim$4.3\%, meaning insufficient room for the larger Se-Se dimer (see Fig.1). Then a compressive strain is introduced and the gap reduction should be largely related to a ``geometrical" effect, unlike the ``electronic" effect for the TiS$_2$Se$^M$.

Before exploring the excitonic properties, it is worth to note that the TiS$_2$Se$^M$ shows a metallic behavior by the PBE. Although the HSE can produce a bandgap, it is computationally unsustainable to solve the BSE on the HSE level at present. Directly applying a scissor correction on a metallic band is problematic. Herein we apply a small in-plane strain to fix this issue. As shown in Figs. 2(e) and 2(f), the gap of 0.08 eV and 0.21 eV is opened under 2$\%$- and 3$\%$-strain at the $\Gamma$ point, respectively. Except for the gap opening, just a slight change is found in comparison with the band structure plotted in Fig. 2(d) which is the one without strain. Below, we will first discuss the excitonic properties of TiS$_2$Se$^M$ by solving the BSE on the 2$\%$-strained electronic structure from the PBE but with a scissor correction to the HSE bandgap of pristine TiS$_2$Se$^M$, namely, 0.58 eV.

\begin{figure}[tbp]
\includegraphics[width=0.95\columnwidth]{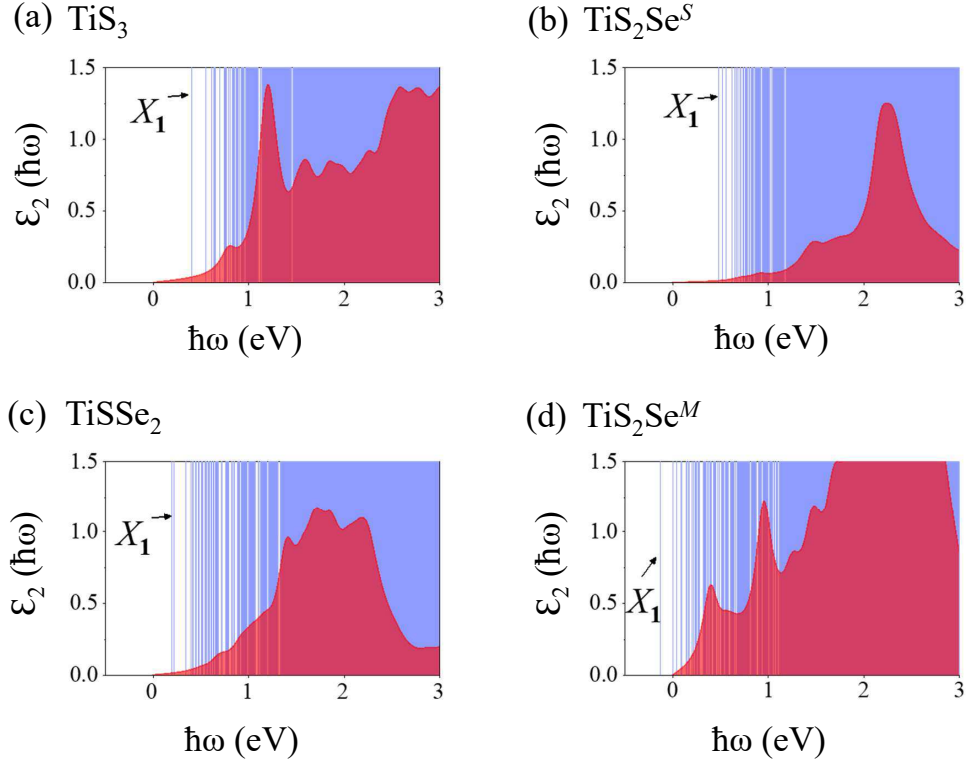}
\caption{\label{fig:fig3} (Color online)  Exciton energies (Blue vertical lines) superimposed on the imaginary part ($\varepsilon_2$) of the BSE dielectric function in the low-energy region for the (a) TiS$_3$, (b) TiS$_2$Se$^S$, (c) TiSSe$_2$, and (d) TiS$_2$Se$^M$. $X_1$ denotes the ground state exciton with the lowest formation energy. A negative energy means spontaneous formation of the corresponding exciton.}
\end{figure}

In Fig. 3 and Table I, we compare the low-energy excitations of the TiS$_3$ and alloys. Because band-edge transitions are dipole-forbidden, the ground state exciton $X_1$ is dark for all the four situations. Forming this exciton costs an energy of 0.4 eV for the pristine TiS$_3$, in agreement with the published literature\cite{usEI}. This value just changes a little to 0.48 eV for the TiS$_2$Se$^S$ and is reduced to half, 0.2 eV, for the TiSSe$_2$. In sharp contrast, it becomes negative, namely, -0.13 eV for the TiS$_2$Se$^M$ as shown in Fig. 3(d), which means spontaneous generation of the $X_1$ exciton. In other words, the one-electron band structure becomes unstable against the transition to an excitonic insulator\cite{Kohn}. Although both the TiS$_2$Se$^S$ and TiSSe$_2$ exhibit the indirect-gap feature, there is only a rather small difference between their direct and indirect gaps. In this regard, that we consider zero-momentum exciton should cause little impact on the real formation energy of the ground state exciton.

It is instructive to analyze in detail the trend that distinct Se substitutions lead to different excitonic properties. By definition, the exciton formation energy is the net difference between exciton binding energy and corresponding one-electron gap. Accordingly, the exciton binding energies are derived and summarized in Table I. Interestingly, they are all in the range 0.71$\sim$0.81 eV. As compared with the notable change of bandgap, the exciton binding energy can be considered almost constant, independent of the concentration and position of Se substitution. Therefore, it is the disproportional effect of Se alloying on the one-electron bandgap and many-body interaction of the system that trigers the excitonic instability in the TiS$_2$Se$^M$.

\begin{figure}[tbp]
\includegraphics[width=0.9\columnwidth]{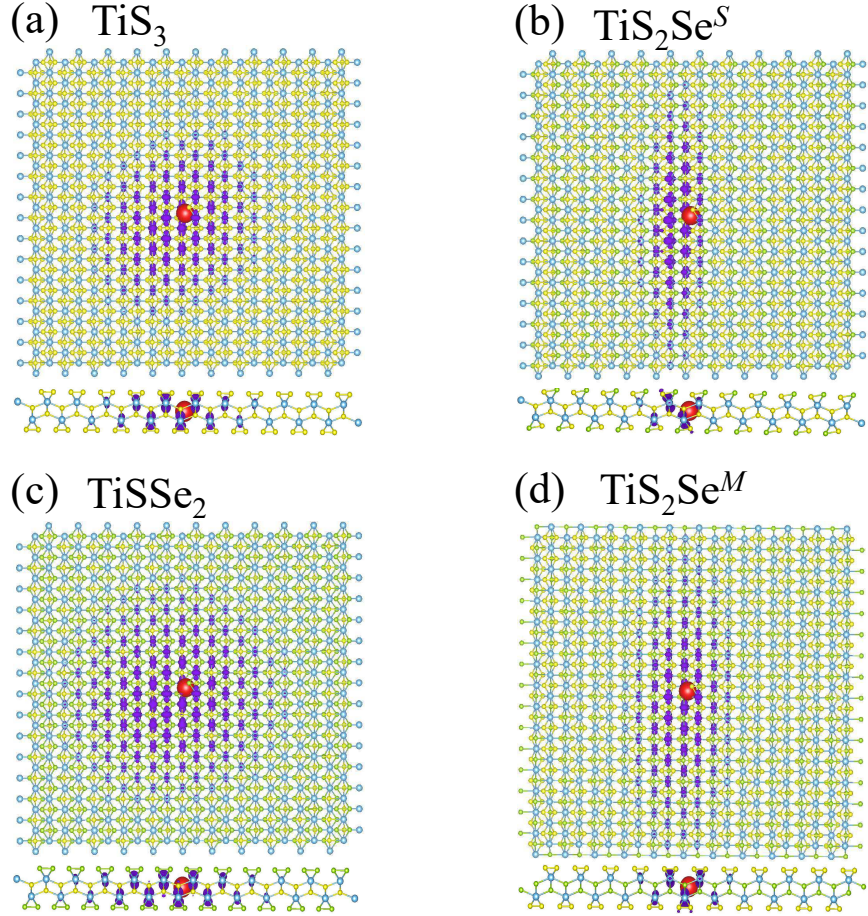}
\caption{\label{fig:fig4} (Color online) Real-space wavefunction plots of the $X_1$ exciton for the (a) TiS$_3$, (b) TiS$_2$Se$^S$, (c) TiSSe$_2$, and (d) TiS$_2$Se$^M$. The hole is fixed at the center (red ball) and the isosurface corresponds to the electron density of 0.06 e/\AA$^3$.}
\end{figure}

To better understand the Se alloying effect on the system electron-hole interaction, we plot the real-space exciton wave-functions for the four cases in Fig. 4. On the appearance, the wave-functions of the TiS$_3$ and the TiSSe$_2$ are similar: the electron is distributed in a circular fashion around the hole; the wave functions of the TiS$_2$Se$^S$ and the TiS$_2$Se$^M$ are similar: the electron is distributed in a quasi-one-dimensional fashion relative to the hole, along the $a_2$ direction. Previous works\cite{Olsen,usPRL} showed the importance of two-dimensional polarizability in determination of electron-hole screening interaction, as well as the binding energy of two-dimensional excitons. Our calculations yield the value of 5.66, 6.44, 6.74, and 6.13 \AA\ for monolayer TiS$_3$, TiS$_2$Se$^S$, TiSSe$_2$ and TiS$_2$Se$^M$. Indeed, the change of two-dimensional polarizability is not large, implying a relatively small change of the system screening effect. However, the trend of its relative size does not correspond to the relative size of exciton binding energy (see Table I). This is probably due to the dark-exciton nature while the reported quantitative relationship between exciton binding energy and two-dimensional polarizability is applicable to the optically-active bright excitons\cite{Olsen,usPRL}.

Next, we discuss the approximation to calculate the excitonic properties of TiS$_2$Se$^M$ in terms of the electronic structure under 2\%-strain. Firstly, comparison between the HSE bands without and with strain [see Figs. 2(d) and 2(e)] reveals little difference. Furthermore, we calculated the electronic structure under 3\%-strain which is shown in Fig. 2(f). It has a direct gap of 0.21/0.94 eV at the PBE/HSE level. On the basis of this PBE band, solving the BSE with a scissor correction to HSE gap (0.94 eV) leads to an exciton formation energy of 0.20 eV. The value is almost identical to the 0.21 eV which is obtained from the PBE band under 2\%-strain coupled with a scissor correction to the gap of 0.94 eV. Almost identical value is indicative of a negligible effect of small strain on the formation energy of the $X_1$ exciton. The corresponding exciton binding energies are also further deduced, namely, 0.74 and 0.73 eV, respectively for the two calculating means. Again, they both fall within the aforementioned 0.71$\sim$0.81 eV range. Note that a similar weak strain-dependence of the exciton binding energy is also found in the monolayer TiS$_3$\cite{usEI}. Altogether, the negative exciton formation energy of -0.13 eV should be quite reasonable for the TiS$_2$Se$^M$, therefore the excitonic instability.

Experimentally, the Se has been successfully alloyed into the TiS$_3$, although on the appearance, our studied cases correspond to a relatively large amount of Se alloying as compared with the current experimental observation\cite{Agarwal}. Two reasons lead us to believe that our model can capture the underlying physics for occurrence of the excitonic instability in the TiS$_{3-3x}$Se$_{3x}$ alloy. On the one hand, the energetics showed that the most stable alloy composition is located at $x$ = 1/3 for the TiS$_{3-3x}$Se$_{3x}$\cite{Agarwal}, which just corresponds to the compounds of here studied TiS$_2$Se$^S$ and TiS$_2$Se$^M$. In practice, the higher amount alloying of the Se might be alternatively obtained by means of other non-equilibrium methods, as will be discussed later on the basis of defect engineering. On the other hand, the effect of Se concentration seems marginal in comparison with that from its substitution position. For instance, we performed the calculations on the case of only one Se substitution for one side S. Now the Se concentration is half that of the TiS$_2$Se$^S$. But the results (not shown here) are essentially similar to those of the TiS$_2$Se$^S$, both in terms of the one-electron band structure and excitonic properties.

In this sense, the location of Se becomes critical to the phase transition from a band-insulator to an excitonic-insulator, i.e., the Se needs to replace the middle S. Here we give some hints by comparing the energetics of different TiS$_{3-x}$Se$_x$ ($x$=0, 1 and 2) alloys. The TiS$_2$Se$^M$ has a higher energy by 74 meV/Se than that of the TiS$_2$Se$^S$, so the Se prefers to replace the side S under a small alloying concentration. This makes sense because there is more space on the edges to accommodate larger Se atoms, consequently lowering the system total energy. Nevertheless, going from TiS$_2$Se$^S$ to TiSSe$_2$ costs an additional energy of 580 meV/Se as compared to going from TiS$_3$ to TiS$_2$Se$^S$. Such an energy requirement is too much as compared to the 74 meV/Se which is the corresponding amount to replace the middle S. Hence, there must be a thermodynamically stable TiS$_{3-x}$Se$_x$ alloy phase in which the S$^M$ has been replaced by the Se before it becomes TiSe$_3$. In addition, given that the Se-to-S$^M$ substitution only costs an additionally moderate energy of 74 meV/Se than the Se-to-S$^S$ substitution, there might be some kind of dynamical control in experiment to solely replace the S$^M$ and directly fabricate the desired TiS$_2$Se$^M$ structure. One possible route is to first create single S vacancy in the TiS$_3$ using methods that have proven to be successful in other transition metal sulfides, such as oxygen plasma exposure\cite{Ye} and  electrochemical desulfurization\cite{Tsai}, and then saturate these S vacancies with the Se atoms. Our calculations show that the Se prefers to bind the S vacancy at the middle rather than at the side as a result of a larger binding energy by $\sim$2 eV. In this way, the Se can be incorporated into the desired middle site.

\begin{figure}[tbp]
\includegraphics[width=0.9\columnwidth]{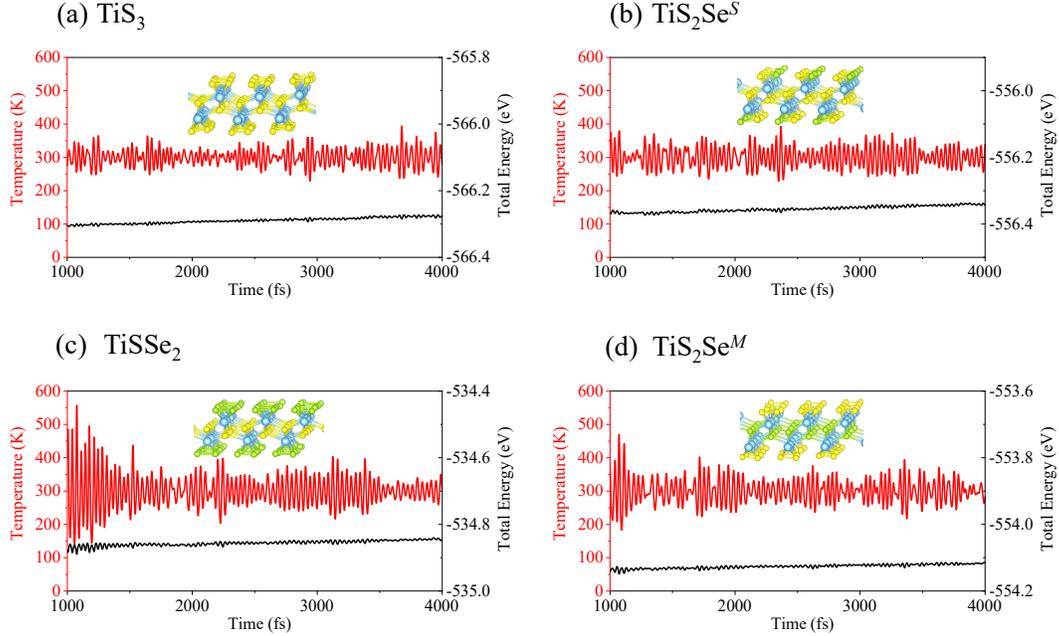}
\caption{\label{fig:fig5} (Color online) Temperature and energy fluctuations with the time step for (a) TiS$_3$, (b) TiS$_2$Se$^S$, (c) TiSSe$_2$, and (d) TiS$_2$Se$^M$. Insets are the corresponding structures at 4 ps.}
\end{figure}

We further test the system stability by molecular dynamic calculations for the four studied systems as implemented in the Vienna ab initio simulation package (VASP)\cite{vasp}. The time step is set to be 1 fs and the simulation lasts 4 ps at 300 K using the Nos\'{e} algorithm. Figure 5 shows the temperature and energy fluctuations for the last 3 ps, which is consistent with the observation that the quintuple-layer structure of the TiS$_{3-x}$Se$_x$ is always preserved in the molecular dynamic simulations. Hence, we can expect that the TiS$_{3-x}$Se$_x$  should be stable with respect to thermal fluctuations up to room temperature.

\vspace{0.3cm}
\textbf{IV. Conclusions}
\vspace{0.3cm}

In summary, the first-principles calculations in combination with Bethe-Salpeter equation reveal the possibility of engineering excitonic insulator via alloying the Se into the TiS$_3$ monolayer. The physics behind lies at the distinct responses of electron and exciton to the Se substitution. While the electronic bandgap can be significantly decreased, the exciton binding energy just varies a little. Replacing the Se for the middle S atoms will lead to spontaneous exciton formation as now the one-electron bandgap has been reduced to less than the exciton binding energy. The energetic calculations demonstrate the feasibility of alloying the Se at the desired middle site. Our work offers a new insight for engineering an excitonic insulator in two-dimensional materials.

\vspace{0.3cm}
\textbf{Acknowledgments}
\vspace{0.3cm}

We acknowledge useful discussions with Zeyu Jiang. This work was supported by the National Natural Science Foundation of China (Grant Nos. 11674071) and the Beijing Institute of Technology Research Fund Program for Young Scholars.

\end{document}